\documentclass{aa}
\usepackage{graphicx}

\begin{document}

   \title{The Self-Enrichment of Galactic Halo Globular Clusters}

   \subtitle{The mass-metallicity relation}

   \author{G. Parmentier \inst{1, 2} 
           \thanks{Visiting Marie Curie Fellow at
           the Institute of Astronomy, Cambridge University }
          \and
           G. Gilmore \inst{2}          
          }

   \offprints{G.~Parmentier (parmentier@astro.ulg.ac.be)}

   \institute{Institute of Astrophysics and Geophysics, 
              University of Li\`ege,
              avenue de Cointe 5, B-4000 Li\`ege  
             \and             
              Institute of Astronomy, Madingley Road, 
              Cambridge CB3 0HA, UK               
             }

   \date{Received ; accepted }

   \abstract{We discuss the existence 
             of a mass-metallicity relation among galactic 
             halo globular clusters.  The lack of any luminosity-
             metallicity correlation in globular cluster systems
             has been used as an argument against
             self-enrichment models of cluster formation.  
             We show that such a 
             relation is statistically present among the galactic Old Halo
             globulars. 
             This observational correlation implies that the 
             least massive old clusters are the most metal-rich.
             This is in contradiction with the idea 
             that, if globular clusters were self-enriched systems,
             the most metal-rich clusters would also be the most massive 
             ones.  We further show that this anti-correlation 
             is as predicted by self-enrichment models. 
   }

\titlerunning{Self-Enrichment of Galactic Halo Globular Clusters}
\authorrunning{Parmentier \& Gilmore}
\maketitle{}

%

\section{Introduction}
Galactic Globular Clusters are among the very first 
bound structures that formed in what later became the Galaxy.  
As such, individual
globular clusters and the Globular Cluster System provide us
with important insights concerning the age, the formation and the 
early evolution of the Galaxy. \\

This paper is the third of a series devoted to the study of a
formation scenario of galactic halo globular clusters, 
namely the self-enrichment 
hypothesis, which develops the Fall \& Rees (\cite{fall}) cluster formation
model.  The model is detailed in 
Parmentier et al.~(\cite{parmentiera})
(hereafter Paper~I) and a summary is provided in Parmentier 
et al.~(\cite{parmentierb}) (hereafter Paper~II), Sect.~2. \\
The model assumes that primordial cold clouds embedded 
in a hot and diffuse protogalactic background (Fall \& Rees \cite{fall}) 
are the gaseous progenitors of galactic halo globular clusters, that is, 
this model assumes baryon assembly predates star formation.
Our model explores the ability of these proto-globular cluster clouds 
to retain the ejecta of a first generation of zero-metal abundance 
stars, born in the central regions of the clouds.  
When the massive stars explode as Type II supernovae, 
they chemically enrich the surrounding pristine interstellar medium
 and trigger the expansion 
of a supershell in which a second generation of nonzero-metal 
abundance stars may form.  The aim
of a self-enrichment scenario is therefore to explain both the 
formation of a globular cluster and the origin of its metal content. \\

One of the key parameters of this class of model is the external 
pressure exerted by the
hot protogalactic background on the proto-clusters.  The higher the pressure is
(i.e. the deeper the proto-cluster is located in the protoGalaxy
in the simplest implementation of the model), the smaller
its mass is, the higher its metallicity will be (see Table~1 of Paper~I). \\
An in-depth discussion of the ensuing Galactic metallicity gradient 
is presented in Paper~II.  We show that, when combined with a pressure profile
scaling as $P_h \propto D^{-2}$, where $P_h$ is the hot protogalactic 
background pressure and $D$ is the galactocentric distance, 
the model is consistent with the 
metallicity gradient observed for the Old Halo 
globular cluster system. \\

There are three aspects of globular cluster formation which self-enrichment
models must specifically address.  The disruptive effects of supernovae, 
and the internal chemical homogeneity are discussed in Paper I.  This paper
considers the third aspect, the extent to which a mass(luminosity)
-metallicity relation is expected and observed.  The Galactic globular 
cluster system is used as a specific example.


\section{Self-enrichment and a mass-metallicity relation}

The lack of any obvious correlation, in any globular cluster system, 
between the 
mass (or the luminosity) and the metallicity of individual globulars is 
often used as an argument against the self-enrichment hypothesis. 
Indeed, were one to assume that gravitational potential gradients
dominated mass loss, the most massive objects would be   
better able to retain their metal-enriched 
supernova ejecta, so that metal abundance should 
increase with cluster mass in case of self-enrichment.
Before adressing the discussion of a luminosity-metallicity relation,
we would like to dismiss this idea that more massive clusters would be more 
metal-rich in the case of self-enrichment.  If a more massive object
is indeed better able to retain more supernova ejecta, 
this larger amount of metallic ejecta is mixed with a larger 
amount of gas. Therefore, no firm conclusion can be drawn concerning the 
resulting metal abundance (or metallicity), i.e. 
the {\sl ratio} of the two increased quantities.  
It is the fractional efficiency of gas retention which is important.
Most importantly, though, mass loss in this class of models is
determined by external gas pressure and not by the pressure equivalent
of the gravitational potential gradient.
This means that the absence
of a mass-metallicity relation, in the sense that the most massive 
globulars would also be the most metal-rich (e.g. McLaughlin 
\cite{mclaughlin}, Barmby et al.~\cite{barmby}), can not be considered as 
evidence against the self-enrichment hypothesis.  
In sharp opposition with these statements, the self-enrichment 
model we develop foresees a mass-metallicity relationship {\sl in 
the sense that the most metal-rich proto-globular clusters are the 
least massive ones}. 

\begin{figure}
\centering
\includegraphics[width=9cm]{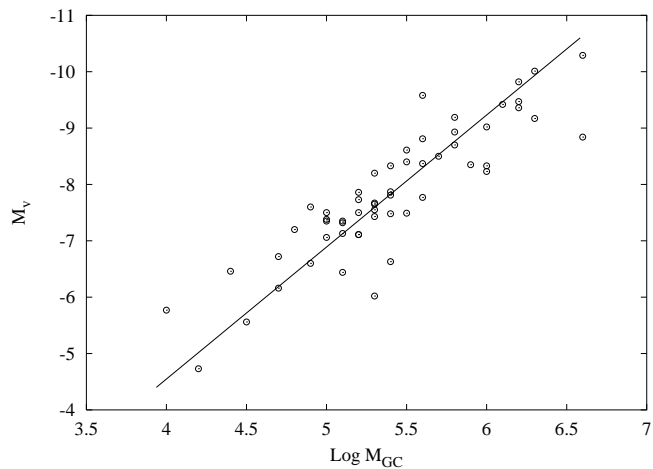}
\caption{Mass-luminosity relation for the 56 globular clusters 
of the Pryor \& Meylan 
(\cite{pryor}) compilation and the corresponding least-squares fit}
\label{MvLogm}
\end{figure}

Unlike globular clusters, 
dwarf galaxies exhibit well-defined correlations between 
luminosity and metallicity (e.g. Gilmore \cite{gilmore}, Mateo \cite{mateob})
such that the dimmest ones are the most metal weak.
The standard explanation for this correlation being self-enrichment 
in the presence of galactic winds which are limited by gravitational 
potential gradients (Dekel \& Silk \cite{dekel}),
Djorgovski \& Meylan (\cite{djorgovski}) 
conclude that globular clusters cannot be self-enriched
systems.  However, Dekel \& Silk (\cite{dekel}) point out that the
dwarf galaxy observed 
luminosity-metallicity relation can be successfully explained only
if the gaseous proto-galaxies are embedded within dominant halos of
dark matter.
While there is indeed clear evidence of the presence of such halos
around dwarf galaxies (Mateo \cite{mateoa}), 
this is not the case for globular clusters (Moore \cite{moore},
Meylan \& Heggie \cite{meylanc}).  Therefore, the Dekel \& Silk model,
built for dwarf galaxies, can certainly be not extrapolated to globular 
clusters.  Moreover, dwarf galaxies and GCs have undergone
very different star formation histories: their respective star formation
rate and duration differ by, at least, an order of magnitude (Gilmore 
\cite{gilmore}).
Dwarf galaxies also exhibit metallicity spreads, often larger than 1\,dex 
(Mateo \cite{mateob}), in marked contrast with the chemical homogeneity of 
globular clusters.
Considering these many differences, the comparison between globular clusters
and dwarf galaxies therefore appears irrelevant.
    
\begin{figure}
\centering
\includegraphics[width=9cm]{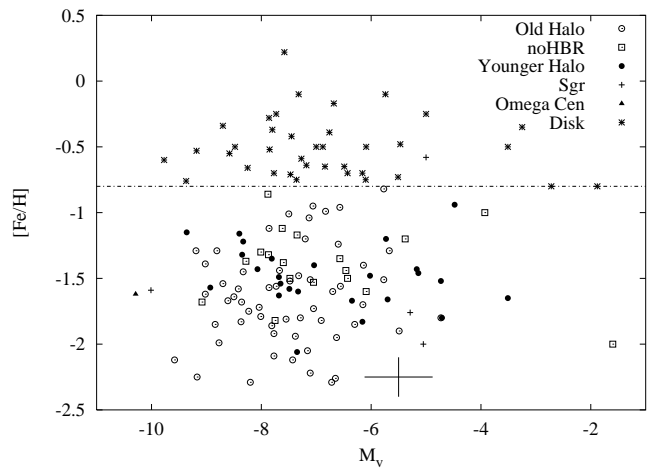}
\caption{Metallicity-luminosity diagram for the whole galactic globular
cluster sytem.  The dashed line 
at [Fe/H]=$-$0.8 represents the generally assumed metallicity limit 
between the halo and the bulge/disk subsystems.  The different types of 
globulars are marked by different symbols.  Old Halo and Younger Halo 
clusters are
respectively represented by open and full circles.  The crosses label
the Sgr globulars.  The open squares (``noHBR'' group) stand for the halo 
globular clusters for which the horizontal brach morphology index
is not given in Harris (1996).  The full triangle and the asterisks 
respectively represent $\omega$ Cen and the bulge/disk clusters}
\label{FeHMv}
\end{figure}

In searching for a luminosity-metallicity relation in the Galactic 
globular cluster system, it should be
kept in mind that, while the observed quantity is the luminosity,
the physical quantity of interest is the mass.  
Figure~\ref{MvLogm} represents the relation between the mass and the absolute
visual magnitude for the 56 globular clusters 
of the Pryor \& Meylan (\cite{pryor})
mass compilation.  The $M_v$ values come from the McMaster Catalogue
(Harris 1996, updated 1999).  
The scatter superimposed on the correlation, of the order of
$\sigma_{M_v}\simeq$0.6, is equivalently the variations 
of the mass-to-light ratio from cluster to cluster. This   
ranges from $\sim$1 to $\sim$4 (Pryor \& Meylan \cite{pryor}) and
reflects possible differences in the initial mass function and the
dynamical evolution of the clusters
\footnote{Part of the scatter also originates
in the uncertainties of globular cluster mass computation, since it relies
on model assumptions such as the isotropy of velocity distibution and a 
maxwellian distribution of velocities.}.  
Therefore, any mass-metallicity correlation will be, at least partly, 
washed out in the corresponding M$_v$-[Fe/H] plot.  This effect is 
illustrated in Fig.~\ref{FeHMv}, the metallicity-luminosity diagram for the
whole globular cluster system
 (the [Fe/H] values are taken from the McMaster Catalogue).
Also plotted are the corresponding error bars in [Fe/H], $\pm$ 0.15\,dex 
(King \cite{king}), and in $M_v$, $\pm$ 0.6\,dex from Fig.~\ref{MvLogm}, 
if the latter is 
considered to be a mass indicator.  The size of the $M_v$ errorbars
(reflecting the different luminosities that a globular cluster 
with a given mass but 
varying mass-to-light ratios may exhibit) 
is clearly not negligeable compared to the size of the observed distribution,
the dispersion of the best-fitting gaussian to the galactic globular cluster
luminosity function 
being $\simeq$1.2 (Harris \cite{harrisa}).
  
Unfortunately, determination of the physical quantity of interest, 
i.e. the relative masses of the globular clusters at their formation, 
is still uncertain 
at least by a factor 2 (Meylan \cite{meylana}).  
For instance, the use of single-mass 
King models is a simplification which tends to underestimate cluster mass 
(Ashman \& Zepf \cite{ashman}, Mermilliod \cite{mermilliod}).  
Therefore, one of the key points in the search for a 
mass-metallicity correlation 
is to use an {\sl homogeneous} set of globular cluster masses
in order to limit additional scatter in the (log$M_{GC}$, [Fe/H]) 
plot.  We use the globular cluster 
mass compilation computed by Pryor \& Meylan (\cite{pryor}):  
this compilation is the most complete set of globular cluster 
masses computed with an internally consistent 
family of multi-component King-Michie models. 

Another source of scatter in the luminosity(mass)-metallicity plot is
introduced by the various origins of the Galactic globulars.
Indeed, evidence has now accumulated that the Galactic globular cluster
system does not consist of globular clusters with a single origin.  
While the majority of globular clusters in the halo are 
old, with a remarkably small age spread (Rosenberg et al.~\cite{rosenberg}),
there is a small subset, particularly among the more metal-rich clusters,
with inferred ages of several Gyr younger than the dominant old
population.  These younger 
globular clusters are either clusters being/having 
been accreted by the Galaxy recently or 
metal-rich clusters associated with the bulge/disk subsystem.  
These clusters being significantly younger, 
their formation is not expected to be 
taken into account by our self-enrichment model, which deals with globular 
clusters whose gaseous progenitors have a primordial composition. \\ 
The age spread highlighted by Rosenberg et al.~(\cite{rosenberg}) 
also confirms the globular cluster system subdivisions
early suggested by Zinn (\cite{zinna}, \cite{zinnb}).
From the point of view of the metallicity distribution, the Galactic 
globular cluster system is composed of two subpopulations, 
a metal-poor halo group and 
a metal-rich, centrally concentrated,  bulge (or disk) group 
(Zinn \cite{zinna}).  Furthermore, the halo subsystem itself includes 
an Old Halo, made of globular clusters perhaps born in situ, during the 
collapse of the protogalactic cloud, and a Younger Halo 
likely made of globulars later stripped from neighbouring dwarf galaxies
(see Paper~II for a review of these evidences).

Since our self-enrichment model deals with globular clusters whose progenitors 
were embedded in the hot phase of the protogalactic cloud and whose
gaseous material was pristine, it is not expected
to apply to the Younger Halo group, 
the presumed accreted component of the halo,
nor to the bulge clusters.  Thus, in what follows, we focus either 
on the coeval and old sample of Rosenberg et al.~(\cite{rosenberg})
or on the Old Halo defined by Zinn.

\section{Comparison of the model with the observations}

\begin{figure}
\centering
\includegraphics[width=9cm]{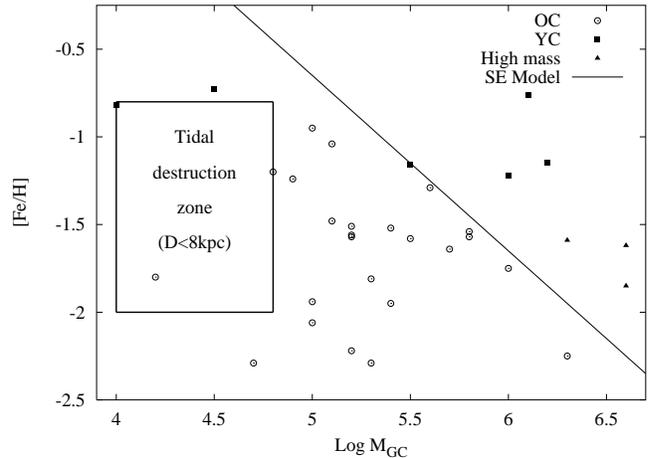}
\caption{[Fe/H] vs log $M_{GC}$ diagram for the globular clusters 
studied by Rosenberg
et al.~(\cite{rosenberg}).  The old (OC) and the younger (YC) globular
clusters exhibit different distributions in the diagram relatively to the
boundary foreseen by the self-enrichment model.  Also shown is the tidal 
destruction zone for the less massive halo globular clusters 
located inside the solar circle}
\label{piotto}
\end{figure}

Our self-enrichment model suggests the existence of an anti-correlation 
between the mass $M$
of a proto-cluster and the metallicity [Fe/H] reached at the end of the 
self-enrichment process, in the sense that the least massive proto-clusters 
create the most metal-rich globular clusters (see Table~1 of Paper~I): 
\begin{equation}
[{\rm Fe/H}]=4.3-{\rm log}~M\;.
\label{logMFeH}
\end{equation}
However, this (anti-)correlation applies to the gaseous progenitors; 
the mass-metallicity relation observed among the studied sample of 
clusters, if any, should only be a relic of Eq.~(\ref{logMFeH}).
For instance, the $-1$ slope will be conserved only if there is a
universal and constant star formation efficiency for the second
stellar generation which forms the majority of the stars from the 
chemically enriched gas swept up from the first generation supernovae.  
Since the mass $M$ of a gaseous progenitor is an upper limit for 
the mass $M_{GC}$ of the globular cluster formed, Eq.~(\ref{logMFeH})
 delimits a permitted area in the
(log$M_{GC}$, [Fe/H]) plot: all the data  
should be located to the left of Eq.~(\ref{logMFeH}) 
(plain curve in Fig.~\ref{piotto} and \ref{LgMFeH0H}).  Figure \ref{piotto}
represents Eq.~(\ref{logMFeH}) together with the globular clusters for which
the age and the mass are respectively provided in Rosenberg et 
al.~(\cite{rosenberg}) and Pryor \& Meylan (\cite{pryor}).  
Obviously, the two Rosenberg et al.~(\cite{rosenberg}) groups (old clusters: 
open symbols; younger clusters: filled symbols) 
behave in a different way compared to our self-enrichment mass-metallicity 
relation.  While the old, coeval and metal-poor GCs are all 
located in the permitted area of the plot, i.e. 
their mass-metallicity diagram is consistent with the self-enrichment of
primordial gaseous progenitors, 
half of the young clusters, either presumed 
accreted or belonging to the bulge subsystem, are located in the 
forbidden area of the diagram, i.e. as expected, 
their formation cannot be accounted for by the self-enrichment model. 
In Fig.~\ref{piotto}, we also represent
three of the most massive globulars (filled triangles), namely 
$\omega$~Cen, M~54 and NGC~5824.  Their location in the forbidden part of the 
plot points to a different star formation history.  This is not
surprising since, at least in the case of $\omega$~Cen and M~54, 
an intrinsic abundance spread is seen.  M~54 is of course a member of the 
Sagittarius dwarf spheroidal galaxy, and is not (yet) a Galactic
globular cluster. \\ 

\begin{figure}
\centering
\includegraphics[width=9cm]{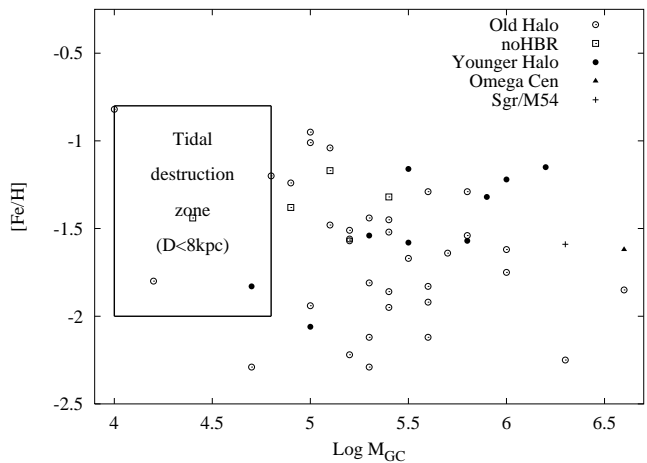}
\caption{[Fe/H] vs log $M_{GC}$ plot including Old Halo and Younger Halo 
subgroups (49 globular clusters).}
\label{LgMFeHWH}
\end{figure}

Figure \ref{piotto} is also clearly depleted in 
low-mass globular clusters (log~$M_{GC} < 4.8$).  
However, at a galactocentric distance
smaller than 8\,kpc, these low-mass clusters are not expected to survive more 
than a Hubble time (see the ``survival triangle'' in the mass vs half-mass
radius diagram defined by Gnedin \&
Ostriker \cite{gnedin}, their Fig.~20a).  
The vast majority of the globular clusters
located at these galactocentric distances, i.e. log~D$<8\,$kpc, 
exhibit a metallicity higher than [Fe/H]=$-$2.  The depletion zone,
represented by the box in Fig.~\ref{piotto} and \ref{LgMFeHWH}, 
is therefore not surprising
and corresponds to the tidal destruction of these low-mass clusters.
The globular clusters used in our Paper are therefore no more 
than a surviving sample.  The distance a given cluster lies to lower masses
from the model upper bound is, to first order, a measure of the star 
formation efficiency of cluster formation.  A ``typical'' surviving cluster
lies a factor of order 5 below the bound, suggesting an efficiency factor
of order 20\%.  As noted above, however, lower mass clusters will have 
preferentially failed to survive until today, so that this value is an 
upper limit.  Star formation efficiencies in the range from a few to
a few tens of percent seem appropriate for most clusters.  Only the few
percent of clusters which are the most massive require star formation 
efficiencies in excess of unity, and so are inconsistent with 
this formation model.  Interestingly, these very massive clusters are 
those which show internal abundance spreads, which are themselves direct
evidence for self-enrichment during cluster formation.  \\

\begin{figure}
\centering
\includegraphics[width=9cm]{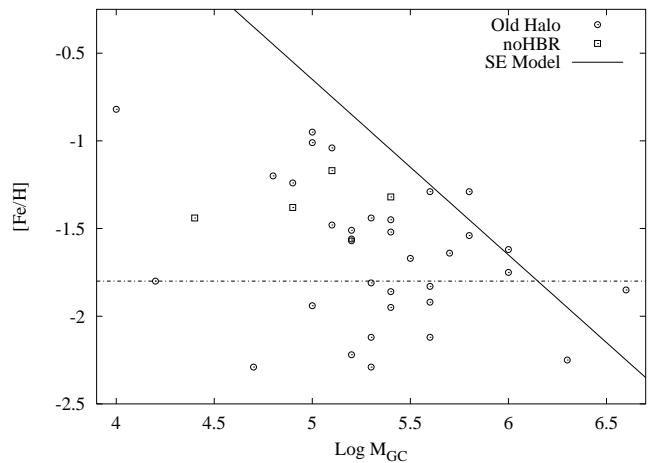}
\caption{[Fe/H] vs log $M_{GC}$ plot for the Old Halo subgroup 
(37 globular clusters).
The self-enrichment model (plain curve) defines a permitted area (left 
part of the plot) in which most of the observational points are located.
The correlation between the globular cluster masses and their metallicities 
is particularly striking for [Fe/H]$> -1.8$}
\label{LgMFeH0H}
\end{figure}

In order to increase our sample and to look for a surviving correlation
between the mass and the metallicity, we also consider the Old Halo subgroup
(Zinn \cite{zinnb}).
As for the metallicity gradient (see Paper~II), an Old Halo/Younger Halo 
separation is fruitful.
Figure~\ref{LgMFeHWH} shows a plot of [Fe/H] versus 
mass for the 49 halo globular clusters whose mass has been computed by 
Pryor \& Meylan (\cite{pryor}): 
there is no correlation between the mass and the metallicity,  
the linear Pearson correlation coefficient being $-$0.15.  \\
Considering the Old Halo group only (Fig.~\ref{LgMFeH0H}), 
as stated in the previous Section,
a weak correlation 
between the logarithm of the mass of the globular clusters and 
their metallicity emerges.
The linear Pearson correlation coefficient improves to a value of 
$-$0.35, with a corresponding probability of correlation of 96.92\%.  
Moreover, most of the Old Halo globulars are located in the 
permitted area of the plot.  
\footnote{If we consider the Old Halo clusters with a 
metallicity larger than $-$1.8,
the increase of the metallicity with decreasing mass is strengthened:  
the linear Pearson correlation coefficient is $-0.64$ corresponding 
to a probability of correlation of 99.93\%. 
One can question about the validity of dividing the Old Halo group
at a metallicity of $-$1.8.  According to Zinn (\cite{zinnc})
and Dinescu et al.~(\cite{dinescu}), the Old Halo group
with [Fe/H]$>-$1.8 is kinematically distinguishable from the Old Halo group
with [Fe/H]$<-$1.8 (their Metal-Poor component).  However, more 
kinematical data are needed to clarify this tentative point.}

\section{Conclusions}
We present the mass-metallicity relation foreseen by our self-enrichment 
model.  At first glance, such a globular cluster formation scenario  
is disproven by the lack of any obvious correlation 
between the luminosity and the metallicity of globular clusters, as claimed 
by previous authors (e.g. Djorgovski \& Meylan \cite{djorgovski},
Ashman \& Zepf \cite{ashman}).
However, we stress here that there are numerous sources of scatter between 
the theoretical (mass, [Fe/H]) relation, applying to the gaseous 
progenitors of globular clusters, 
and the observed (luminosity, [Fe/H]) plot, applying
to globular clusters.  These scatter sources are, for instance, 
the star formation efficiency with which the globular cluster stars form out of
the proto-globular cluster cloud, the mass losses undergone by globular
clusters with time (see the tidal tails exhibited by some clusters), 
the variations in the mass-to-light ratio from one globular
cluster to another.    
We also caution that the search for a mass-metallicity correlation should be 
restricted to a given globular cluster subpopulation, namely the Old Halo 
group.
Despite the numerous sources of scatter,  
the globulars of this group are characterized by a boundary in the 
mass-metallicity diagram and by a correlation in the sense 
expected by simple pressure-bounded self-enrichment models. 
Self-enrichment models remain a viable 
hypothesis for galactic halo globular cluster formation.

\begin{acknowledgements}
This research was supported partly by the European Community under grant
HPMT-CT-2000-00132 and partly by contracts P\^ole d'Attraction 
Interuniversitaire P4/05 (SSTC, Belgium) and
FRFC F6/15-OL-F63 (FNRS, Belgium).      
\end{acknowledgements}


\begin{thebibliography}{}
\bibitem[1998]{ashman}
Ashman, K.M., \& Zepf S.E. 1998, Globular Cluster Systems, Cambridge
Astrophysics Series
\bibitem[2000]{barmby} 
Barmby, P., Huchra, J.P., Brodie, J.P., Forbes, D.A., Schroder, L.L., \&
Grillmair, C.J. 2000, AJ, 119, 727
\bibitem[1986]{dekel}
Dekel, A. \& Silk, J. 1986, ApJ, 303, 39
\bibitem[1999]{dinescu}
Dinescu, D.I., Girard, T.M., \& Van Altena, W.F. 1999, AJ, 117, 1792 
\bibitem[1994]{djorgovski} 
Djorgovski, S., \& Meylan, G. 1994, AJ, 108, 1292 
\bibitem[1985]{fall} 
Fall, S. M., \& Rees, M.J. 1985, ApJ, 298, 18
\bibitem[2000]{gilmore}
Gilmore, G. 2000, in ASP Conf. Ser. 230, Galaxy Disks and Disk Galaxies, 
eds. J.G. Funes \& E.M. Corsini, 3
\bibitem[1996]{gnedin}
Gnedin, O.Y., Ostriker, J.P. 1997, ApJ, 474, 223
\bibitem[1991]{harrisa}
Harris, W.E. 1991, ARA\&A, 29, 543
\bibitem[1996]{harris} 
Harris, W.E. 1996, AJ, 112, 1487 
\bibitem[1999]{king}
King, I.R. 1999, In: Martinez Roger C., P\'erez Fournon I., Sanchez F. (eds) 
Cambridge University Press, Globular Clusters, 1
\bibitem[1996]{mateoa}
Mateo, M. 1996, in ASP Conf. Ser. 92, Formation of the Galactic Halo ...
Inside and Out, eds. H. Morrison \& A. Sarajedini, 434
\bibitem[2000]{mateob}
Mateo, M. 2000, in The First Stars (Proceedings of the MPA/ESO Workshop 
held at Garching, Germany, 4-6 August 1999), eds. A. Weiss, T.G. Abel \& 
V. Hill, Springer, 283
\bibitem[1997]{mclaughlin}
McLaughlin, D.E. 1997, PhD Thesis, McMaster University (Canada)
\bibitem[2000]{mermilliod}
Mermilliod, J.-C. 2000, in ASP Conf. Ser. Vol. 211, 
Massive Stellar Clusters,
eds. A.~Lancon \& C.M.~Boily (San Fransisco: ASP), 43 
\bibitem[2000]{meylana}
Meylan, G., 2000a, In: A. Noels, P. Magain, D. Caro, 
E. Jehin, G. Parmentier, A. Thoul (eds.) 35$^{th}$ Li\`ege 
International Astrophysics Colloquium, The galactic halo: 
from globular clusters to field stars, p~543 
\bibitem[1997]{meylanc}
Meylan, G., \& Heggie, D.C. 1997, A\&AR, 8, 1 
\bibitem[1996]{moore}
Moore, B. 1996, ApJ, 461, L13
\bibitem[1999]{parmentiera} 
Parmentier, G., Jehin, E., Magain, P., Neuforge, C., Noels, A., \&
Thoul, A. 1999, A\&A, 352, 138
\bibitem[2000]{parmentierb} 
Parmentier, G., Jehin, E., Magain, P., Noels, A., \&
Thoul, A. 2000, A\&A, 363, 526
\bibitem[1993]{pryor} 
Pryor, T., \& Meylan, G. 1993 in ASP Conf. Ser. Vol. 50, 
Structure and Dynamics of Globular Clusters, eds. S.G. Djorgovski \& G. Meylan
(San Fransisco: ASP), 357 
\bibitem[1999]{rosenberg}
Rosenberg, A., Saviane, I., Piotto, G., Aparicio, A. 1999, AJ, 118, 2306
\bibitem[1985]{zinna}
Zinn, R. 1985, ApJ, 293, 424
\bibitem[1993]{zinnb} 
Zinn, R. 1993, in ASP Conf. Ser. Vol. 48, 
The Globular Cluster-Galaxy Connection,eds. G.H. Smith \& J.P. Brodie
(San Fransisco: ASP), 38 
\bibitem[1996]{zinnc}
Zinn, R. 1996, in ASP Conf. Ser. Vol. 92, 
Formation of the Galactic Halo...Inside and Out,
eds. H.~Morrison \& A.~Sarajedini (San Fransisco: ASP), 38 
\end{thebibliography}
\end{document}